# Unraveling Extreme Weather Impacts on Air Transportation and Passenger Delays using Location-based Data

*Chia-Wei Hsu, Chenyue Liu, Zhewei Liu\*, Ali Mostafavi*

Urban Resilience.AI Lab Zachry Department of Civil & Environmental Engineering, 3136 TAMU, College Station, TX
\*Corresponding author
E-mail address: lzwgre@gmail.com

## Abstract

Extreme weather poses significant threats to air transportation systems, causing flight rerouting and cancellations, as well as passenger travel delays. With the growing frequency of extreme weather hazards, it is essential to understand the extent to which disruptions in flights and subsequent cancellations impact passenger delays. This study focuses on quantifying the impacts of a recent extreme weather event (2022 Winter Storm Elliott) on the U.S. air transportation system by investigating passenger delays measured based on dwell time at airports using privacy-preserving location-based datasets. The study determines total dwell time and dwell time per anonymized user at airports during the extreme weather event and computes the impact based on changes in values compared to the same period in the previous year. The results show that the storm event caused passengers significant delays, as characterized by a substantial increase in airport dwell time. Factor analysis shows that airports with a greater passenger flow and a greater portion of flights from decentralized airlines aggravated passengers delays during the winter storm. The vulnerability of airports was mainly due to the direct storm exposure, and the influence of network cascading impacts were limited. The findings of this study provide novel insights and quantification of the extent of extreme weather impacts on air transportation at individual airports and national levels. These outcomes could inform airport owners and operators, as well as airlines, about the extent of vulnerability and provide useful information for weather-related risk assessment of air transportation systems.

# 1. Introduction

The growing frequency of extreme events, such as winter storms, hurricanes, and extreme temperatures, disrupts the regular operation of the air transportation systems, causing travel delays and subsequent economic costs to travelers, airports, and airline careers (Doll, Klug et al., 2014). Despite the increasing awareness of the impacts of extreme weather events on air transportation, the extent of impacts on passenger travel delays is not fully understood nor quantified. It is estimated that the hazards of extreme weather events will be further aggravated due to the greater likelihood and severity of events (IPCC, 2018) and consequential impacts on air transportation. Thus, it is critical to better understand and quantify the impacts of extreme weather events on air transportation systems, especially from the perspective of passenger delays.

Previous studies have pinpointed the primary climate events affecting the airports' infrastructure and operations, including sea-level rise and urban flooding, extreme temperatures, intense storms, and heavy precipitation. (ICAO, 2021). These extreme weather events can take place individually and jointly and give rise to complicated, intertwined, and ripple effects on the operation and security of airport systems. For example, the accompanying strong winds of intense storms could threaten the aircraft safety during takeoff and landing, and may also require runway closure in extreme cases, causing substantial flight delays and cancellations. Strong storms usually bring about heavy precipitation within a short period of time, which in turn could overwhelm airports' drainage systems (Voskaki, Budd et al., 2023). In 2013, a severe winter storm combined with heavy precipitation at London's Gatwick Airport caused delays and disruption for more than 16,000 travelers during the Christmas holiday season (Agency, 2016). Abnormal temperatures (extreme heatwaves and cold snaps) are another significant threat to airport infrastructures brought by climate change. Very high and low temperatures can cause equipment failure and improper operations. Also, extreme temperatures can affect runway performance. Very high temperatures can cause the runway to be excessively soft, making an unstable base for aircraft landing (De Vivo, Ellena et al., 2021). For example, in July 2018, prolonged high temperatures caused serious damage to the north runway of Hanover Airport in Germany and resulted in dozens of flight cancellations. Very low temperature may increase the runway's slipperiness and undermine the planes' braking effectiveness (Nuijten, 2016). Another effect of extreme temperatures is the planes' maximum takeoff weights, which are affected by the changing air density due to humidity and temperature fluctuations and are especially influential for airports with higher altitudes and shorter runways (Zhao, 2020).

The increased frequency and intensity of extreme weather events highlights the importance of understanding the impacts of these events on air transportation systems and subsequent economic and social consequences. Most existing studies (such as Chen and Wang, 2019; Zhou and Chen, 2020) focused primarily on examining the impacts of extreme weather events on air transportation systems based on analyzing the extent of flight cancellations and flight delays. Such impacts on the flight delays are typically characterized as the time difference between the scheduled and actual flight departure/arrival (Borsky and Unterberger, 2019). Flight cancellations and flight delays, however, do not fully capture the extent of effects on passenger travel delays. This limitation can be effectively addressed using location-based data.

The emergence of location-based services has given rise to novel risk evaluation approaches based on fine-grained human mobility datasets (Liu et al., 2021; Coleman et al., 2022; Zhang and Li, 2022). Such datasets can provide high spatiotemporal resolution insights regarding population activity and mobility that can be used to understand how people move, act, and respond in hazard scenarios (Wang et al., 2020; Lee et al., 2022; Liu et al., 2022; Rajput et al.,

2022; Rajput et al., 2022). By evaluating the changes in population activity and mobility patterns, the extent of impacts of hazard-induced disruptions can be effectively evaluated (Hsu et al., 2022; Lee, et al., 2022; Rajput and Mostafavi, 2022; Wang and Taylor, 2016; Liu et al., 2019). While location-based data has been used for evaluating hazard-induced perturbations in different infrastructures (Lee et al., 2022), the use of these high-resolution data has been rather limited in examining the vulnerability of air transportation systems to extreme weather events. This study addresses this gap and uses fine-grained location-based data for examining the effects of a recent extreme weather event, 2022 Winter Storm Elliott, on the U.S. air transportation system by quantifying and comparing the passenger delay (derived from passenger dwell time at airports).

Our dataset captured 41,344,261 anonymized trips at 62 major U.S. airports and records their dwell time during the steady-state period (December 21–26, 2021) and perturbation period (December 21–26, 2022). By comparing changes in passenger dwell times across different regions and time periods, this study aims to answer the following specific questions: (1) the extent to which the extreme weather events disrupt the air transportation system and which are the most disrupted airports from a passenger-delay perspective; (2) the extent to which airports are affected by direct hazard exposure versus cascading impacts through network effects; (3) the extent to which the airlines' operating models (decentralized versus centralized) affects the vulnerability of airports. Figure 1 shows the framework for this study, including the data used, metrics computed, and the types of analysis performed to address our research questions. The details related to the dataset and methods are discussed in the next section.

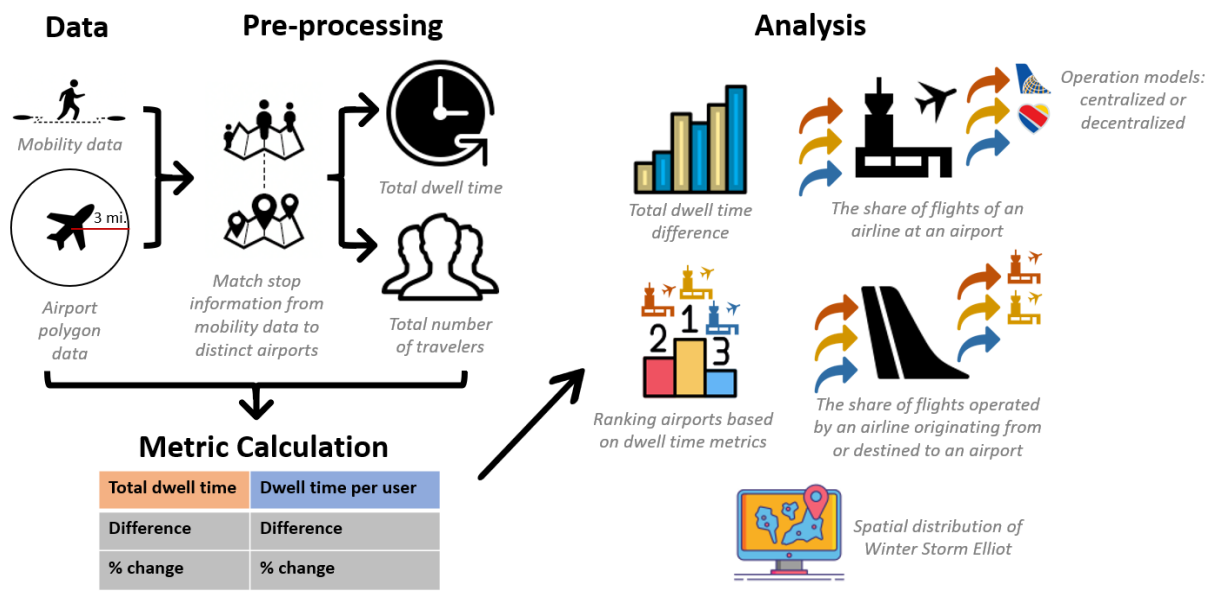

**Figure 1**. Overview of research framework: Location-based data and airport polygon data are used to estimate passengers dwell time at airports; then the impacts are evaluated based on the difference and percentage change of total dwell time and dwell time per user. Based on the total dwell time difference, vulnerable airports are ranked. The effects of operation model of airlines and direct exposure to Winter Storm Elliot are examined.

## 2. Study Context

This study collected data from 2022 Winter Storm Elliott, an extreme winter storm event in North America. From December 21 through 26, 2022, the event produced strong winds, snowfall, and record-low temperatures in the majority of the United States. Due to the extreme weather, more than 18,200 flights (December 2022 North American winter storm, 2023) were canceled in the United States between December 22 and 28, 2022, causing severe travel delays for passengers and seriously disrupting the operations of U.S. aviation system. Figure 2 shows the impact of Winter Storm Elliot across the continental United States on December 22, 2022, provided by the National Weather Service (NWS). Different colors denote different types of warnings. The population threatened by the different extent of impact is recorded.

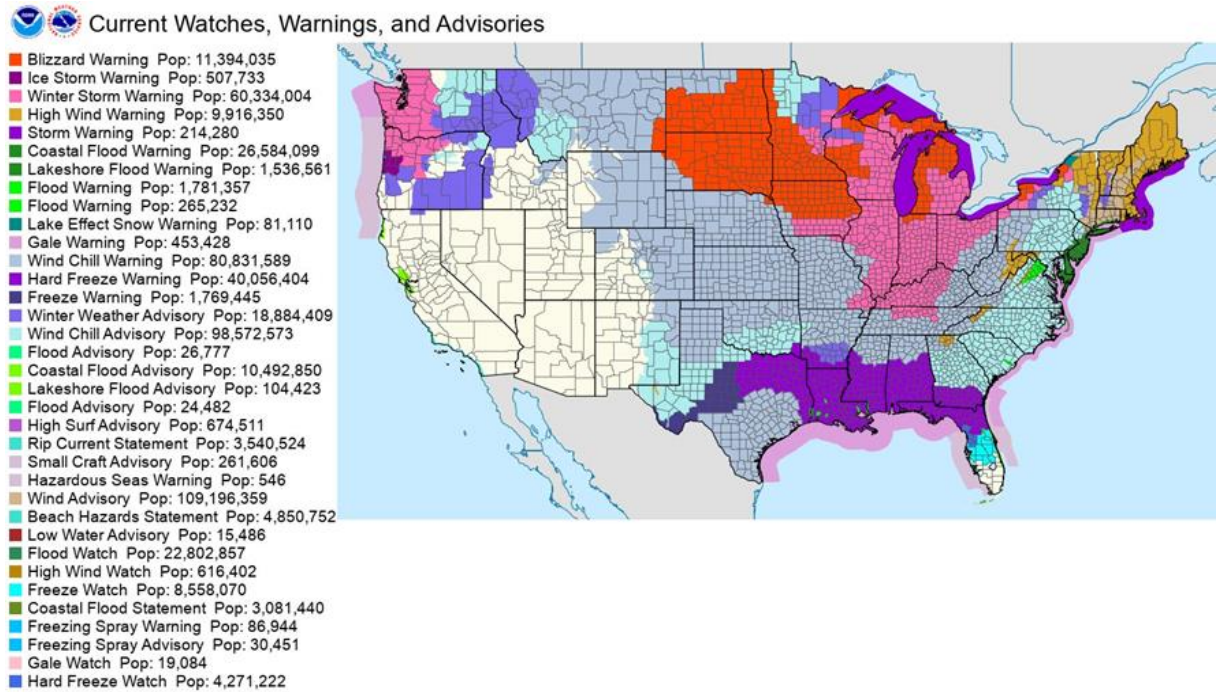

**Figure 2.** The impact of Winter Storm Elliot across the United States on December 22, 2022. (Source: National Weather Service)

## 3. Data Collection and Processing

### 3.1. Airports Information

This study focuses on airports located within the geographical boundaries of the United States. The airport location data utilized in this research was obtained from a publicly accessible website, ourairports.com. The database incorporates comprehensive information on 11,425 airports across the United States. The airports are categorized into diverse types, such as large airports, medium airports, small airports, heliports, balloon ports, and seaplane bases. The scope of this research is limited to large and medium airports, resulting in the selection of 873 airports for further analysis. For each of these airports, we acquired data about the airport's name and the centroid point, which was subsequently used to create a polygon enveloping a three-mile buffer zone around each airport to ensure maximal coverage.

Moreover, additional data related to airport operations were collected from the Bureau of Transportation Statistics, which includes origin and destination airport identification numbers (ID) data. The airport ID is a distinctive numerical code assigned by the U.S. Department of Transportation (DOT) to differentiate between airports.

### 3.2. Dwell time estimation

Airport dwell time was estimated based on the location-based data obtained from the Spectus

Data Cleanroom, a location intelligence platform that collects and processes de-identified data from users who have consented to share their location data for research purposes via a GDPR and CCPA compliant framework. With the high accuracy of GPS location, Spectus calculates people's moving trajectories of devices. On a single day, an average of 100 data points per device may be obtained. Each recorded coordinate is saved as a tuple of longitude and latitude. It is worth noting that all information is de-identified; that is, no personal information of the individual is recorded. Spectus has assigned a hashed and encrypted ID to each device, while also taking additional steps to preserve privacy such as obfuscating home locations to the census tract level, and discarding data from sensitive points of interest. In this study, we use the clustered stop data, which contains information on the stop-by date, time, stop location, and the length of time the device stayed at that point (dwell time). We have used the data collected from December 21, 2022 to December 26, 2022 as the affected period, and the same period in the previous year as the base time period. By combining the stop table with the airport polygons, we determined the number of devices that stayed at each airport and the total amount of time they spent at each airport per day during both the steady-state (December 2021) and perturbation (December 2022) periods.

## 4. Methodology

We used total dwell time and dwell time per use as the main metrics for evaluating the impacts of the Winter Storm on the air transportation system. Dwell time data is used to assess the extent of impact at each airport while flight data provides information about routes and airline operators at airports.

First, we examine the overall impact of Winter Storm Elliot from December 21 through December 26, 2022. We quantify the extent of impact of the storm's impact by comparing the difference in total dwell time and percentage change with the same period in 2021 for each airport. The aggregated total dwell time is obtained by first summing daily total dwell time at airports then summing up across the total days within the analysis period. Second, to understand which airports were more vulnerable to the winter storm's impact, we examined the dwell time at each airport. For each airport, we calculated the total dwell time and dwell time per user in each day within the perturbation period and compared it to the steady-state period. Then, we ranked the airports based on the increase in maximum total dwell time and dwell time per user change.

The third step of the analysis focused on understanding the extent to which impacts can be attributed to: (1) centralized or decentralized airline operation models, and (2) direct hazard exposure versus network effects. We used the flight data during the steady-state period to obtain the proportion of flights by different airlines at each airport. We calculated the share of flights of an airline at an airport and also the share of flights operated by an airline originating from or destined for an airport. The share of flights for an airline at an airport or the composition of the flights at an airport indicates how the overall performance of an airport can be attributed to the operation models of the airlines. The share of flights operated by an airline originating from or destined for an airport captures how an airline operates, whether it has a centralized or decentralized operation model. The centralized model has significant hubs, which means that a large proportion of flights originate from or are destined to a limited number of airports. In contrast, the decentralized model does not have significant hubs, and the quantity of flights originating from or destined to different airports does not differ significantly. To compare whether the number of flights differs significantly or not, we calculate the standard deviation of the number of flights originating from or destined for different airports. If the standard deviation is large, it suggests that the number of flights differs significantly, indicating a centralized model. On the other hand, if the standard deviation is small, it indicates a small different in the number of flights, suggesting a decentralized model. This examination logic helps to distinguish between the two airline operation models. Finally, based on the spatial distribution of changes in dwell time and dwell time per user in each airport, we investigate whether the impacts were spatially co-located with areas with direct exposure to the winter storm.

## 5. Results

### 5.1. Overall impact

First, we examine the overall impact of the winter storm from December 21 through December 26, 2022. The comparison between the same period in 2021 and 2022 is shown in figure 3. The aggregated total dwell time is 61,585,574 hours and 22 minutes for 2021 and 73,556,239 hours and 26 minutes for 2022, represented by blue bars and orange bars, respectively, in figure 3. There is a 11,970,665 hour and 4 minute increase in the aggregated total dwell time due to the winter storm. Considering the percentage of the population for which Spectus data provides a sample, the actual impacts in terms of total dwell increase time would be multiple times more. For example, assuming Spectus data captures 15% to 25% of the total population, the total increase in airport dwell time would be between 44 million to 74 million hours. According to the U.S. Department of Transportation, value of travel time savings (VTTS) for personal travel by air is estimated to be $36.1 per hour (White, 2016). The winter storm could cause a $1.6 billion to $2.7 billion loss.

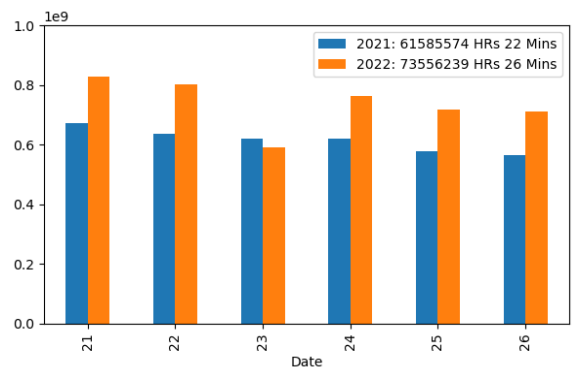

**Figure 3**. The aggregated total dwell time from December 21 through December 26, 2021is 61,585,574 hours in 2021 (blue), representing a steady-state period, and 73,556,239 hours in 2022 (orange), represented a perturbed period. An increase of 11,970,665 hours is due to the winter storm.

### 5.2. Vulnerability of airports

For each airport, the differences and percentage change in total dwell time and dwell time per user were computed for each day. In Figure 4, bars with positive values mean the values of total dwell time and dwell time per user increased and bars with negative value mean the metrics' value decreased. Figure 4 also shows the ranking of airports based on the maximum difference in total dwell time between the perturbation period and the steady-state period. This metric represents the overall vulnerability of an airport by capturing whether its total passenger dwell times increased during Winter Storm Elliot. Chicago Midway International Airport, La Guardia Airport, Reno Tahoe International Airport, Harry Reid International Airport, and Phoenix Sky Harbor Airport were the airports most affected by Winter Storm Elliot in terms of the difference in total dwell time. Figure 5 shows the ranking of airports based on the maximum difference in dwell time per user between the perturbation period and the steady-state period. This metric represents passenger delays due to additional time spent at an airport, capturing if their average dwell time increased during Winter Storm Elliot. Buffalo Niagara International Airport, Portland International Airport, Seattle-Tacoma International Airport, Cleveland Hopkins International Airport and General Mitchell International Airport were the airports most affected by Winter Storm Elliot in terms of the difference in dwell time per user.

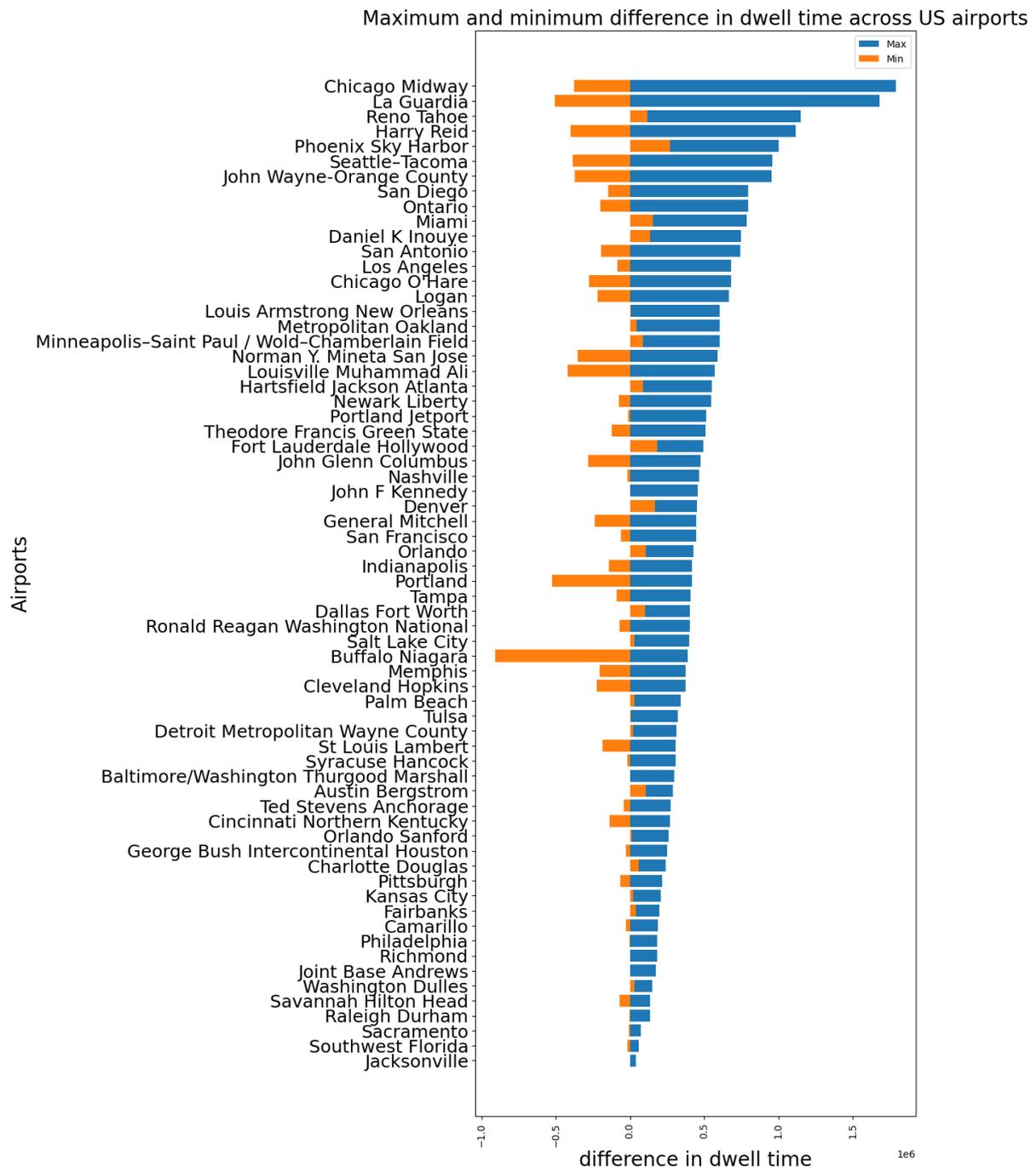

**Figure 4**. Ranking of airports based on the maximum difference in total dwell time between the perturbation period and the steady-state period. The blue bars represent the maximum difference; the orange bars represent the minimum difference in total dwell time. Higher ranking indicates an airport is more affected by the winter storm.

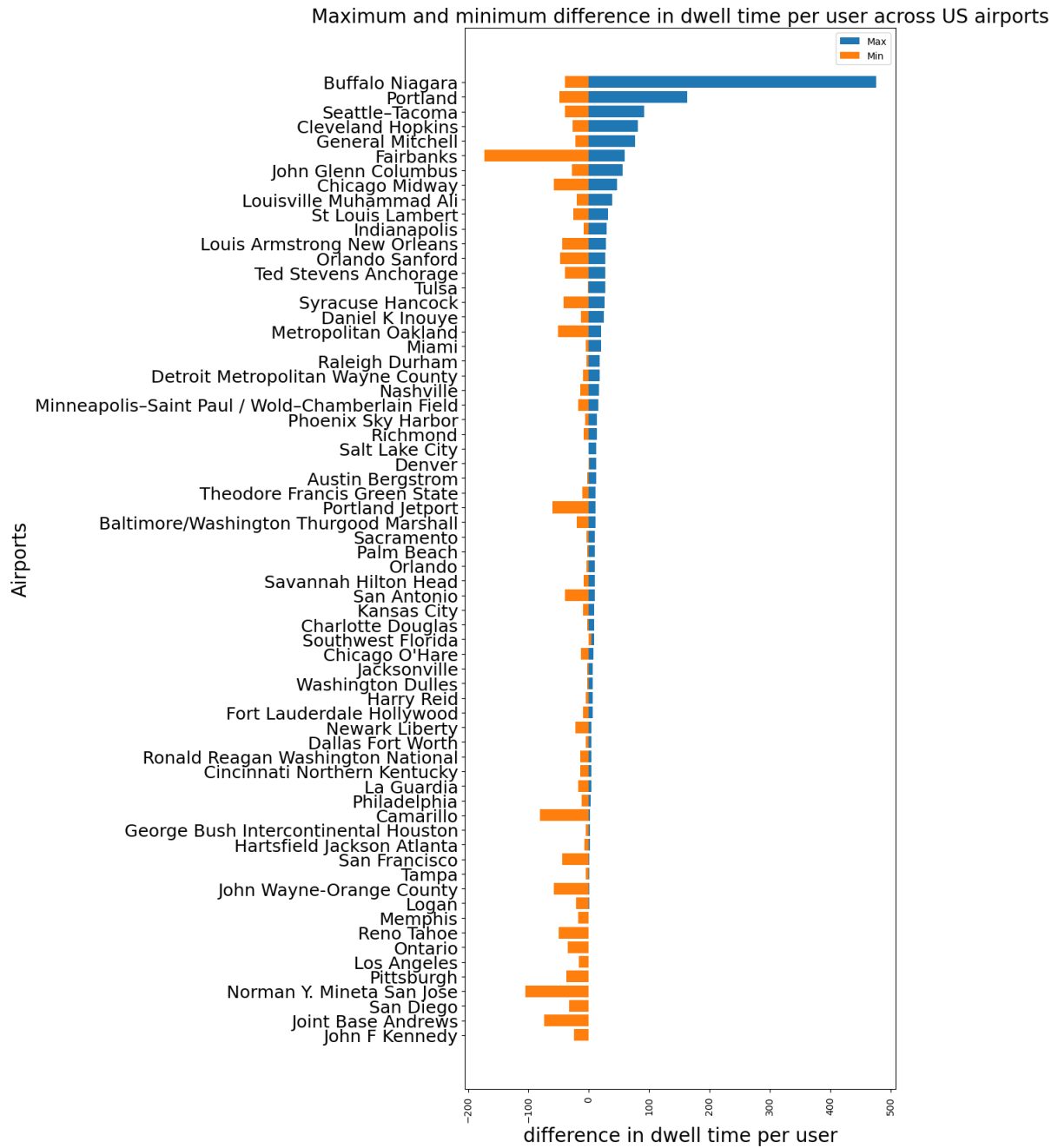

**Figure 5**. Ranking of airports based on the maximum difference in dwell time per user between the perturbation period and the steady-state period. The blue bars represent the maximum difference; the orange bars represent the minimum difference in total dwell time. Higher ranking indicates an airport being more affected by Winter Storm Elliot.

Figure 6 shows the ranking of airports based on the maximum percentage change in total dwell time between the perturbation period and the steady-state period. This metric represents the overall vulnerability of an airport, capturing whether its total dwell times increased relative to 2021 during Winter Storm Elliot in 2022. Denver International Airport, Orlando International Airport, Fairbanks International airport, Kansas City International Airport and Nashville International Airport were the airports most affected by Winter Storm Elliot in terms of the percentage change in total dwell time. These are large airports that have larger passenger flows. Figure 7 shows the ranking of airports based on the maximum percentage change in dwell time per user between the perturbation period and the steady-state period. This metric represents passenger delays due to relative additional time spent at an airport, capturing whether their average dwell time increased during Winter Storm Elliot. Similar to the results for the difference between dwell time per user, Buffalo Niagara International Airport, Portland International Airport, Seattle-Tacoma International Airport, John Glenn International Airport, and General Mitchell International Airport were the airports most affected by Winter Storm Elliot in terms of the percentage change in dwell time per user.

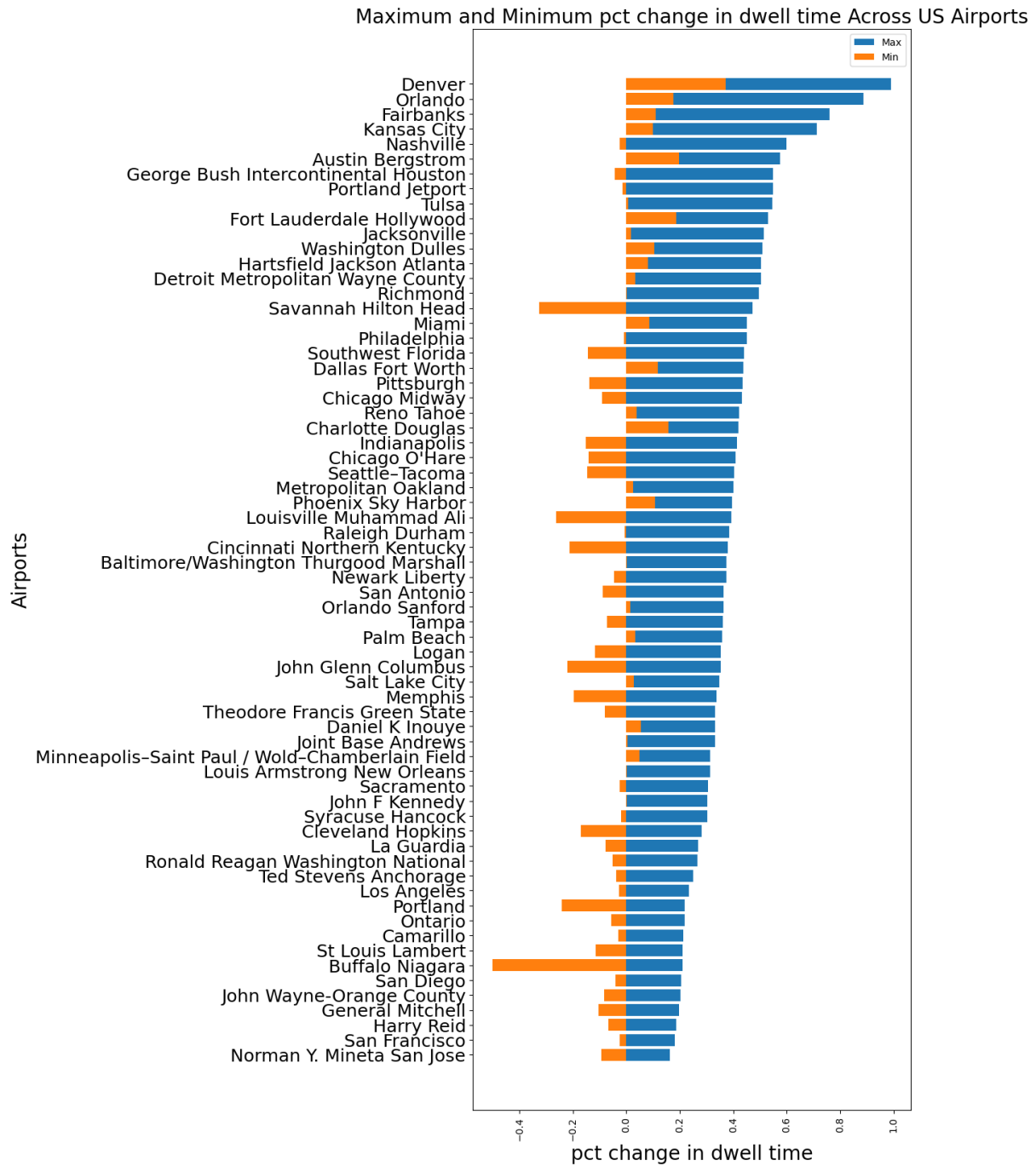

**Figure 6**. Ranking of airports based on the maximum percentage change in total dwell time between the perturbation period and the steady-state period. The blue bars represent the maximum percentage change; the orange bars represent the minimum percentage change in total dwell time. Higher ranking indicates an airport being more affected by Winter Storm Elliot.

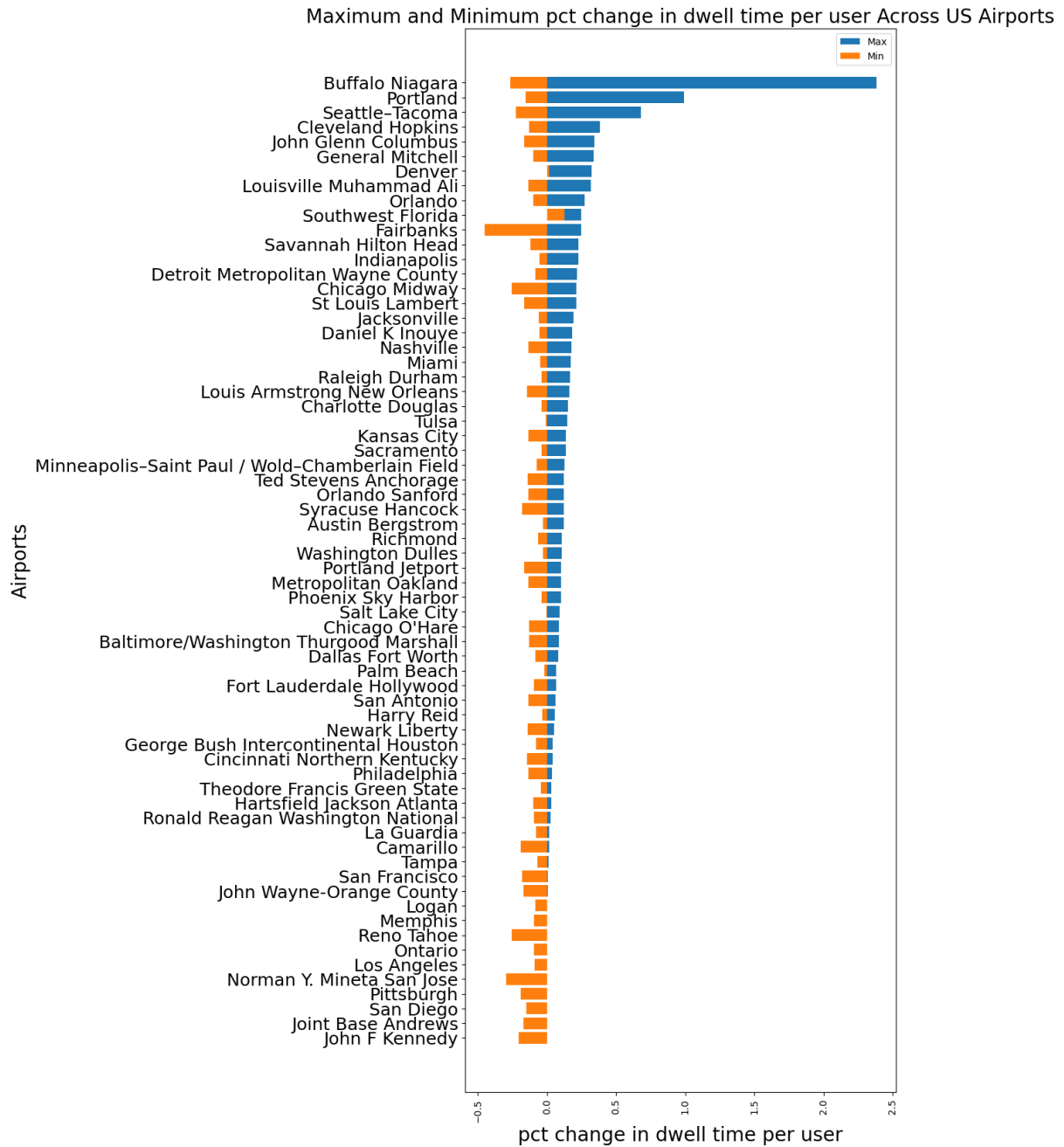

**Figure 7**. Ranking of airports based on the maximum percentage change on dwell time per user between the perturbation period and the steady state period. The blue bars represent the maximum percentage change; the orange bars represent the minimum percentage change on total dwell time. Higher ranking indicates an airport being more affected by Winter Storm Elliot.

From the above results, dwell time per user may be a more informative metric in measuring the impact of the event, as the results related to difference and percentage change are similar. Also, it should be noted that the value of the total dwell time metric is generally proportional to the increase in number of users during this period. Thus, the total dwell time metric may indicate how the passenger flows changed during this period, while the dwell time per user metric better reflects the direct impact on individual passengers.

## 5.3. Relationship of impacts with the operation models of airlines

Based on the previous findings, the next analyses focus on the percentage change of dwell time per user metric. By examining the highest-ranked airport—Buffalo Niagara International Airport, based on the change of dwell time per user—Figure 8(a) decomposes the share of flights of each airline at this airport. The result shows that Midwest Airlines (36.2%) and Southwest Airlines (35.7%) have the highest share, 71.9% combined. Figure 8(b) shows that these two airlines ranked low regarding the standard deviation of the shares, meaning that their operation model is more decentralized. Similar results can be found in most other airports ranked high based on the change in dwell time per user as shown in Figure 9. For example, Portland International Airport, Seattle-Tacoma International Airport, Cleveland Hopkins International Airport and John Glenn Port Columbus International Airport have a high percentage of their flights from airlines that are more decentralized. Thus, the results reveal that the greater extent of impact from the winter storm (based on dwell time per use increase) can be attributed to airlines whose operation model is more decentralized. It should be noted that Seattle-Tacoma International Airport also ranked high, but its flight share differs from the other higher-ranking airports. A large proportion of flights are operated by Alaska Airlines, an airline that has a centralized operation model because it has a strong connection with airports in Alaska. There may not be many other airport options for diverting the flights under winter storms, thus causing unavoidable delays and cancellations.

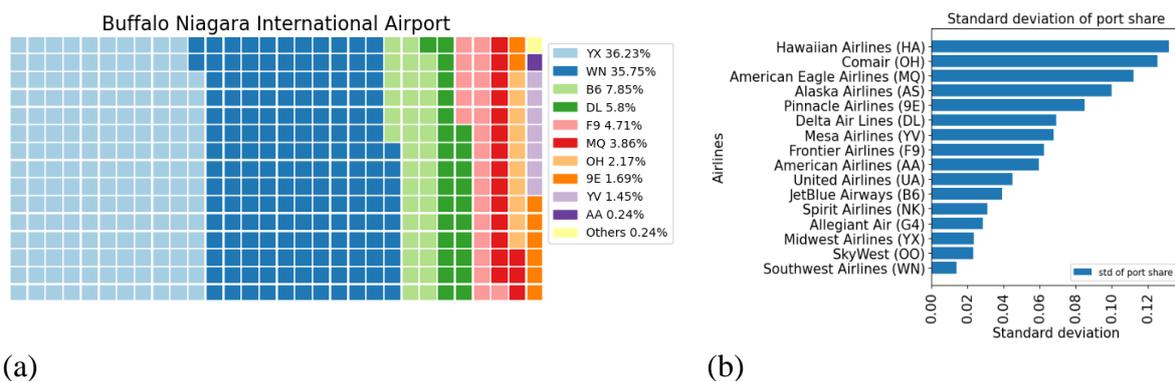

(a) (b)

**Figure 8**. (a) The share of flights of an airline at Buffalo Niagara International airport; (b) airline rankings in terms of the standard deviation of their flight shares; higher rankings indicate it is applying an operating model which is more centralized. A high proportion (71.6%) of flights at Buffalo Niagara International Airport are operated by airlines (Midwest, Southwest) applying a decentralized model.

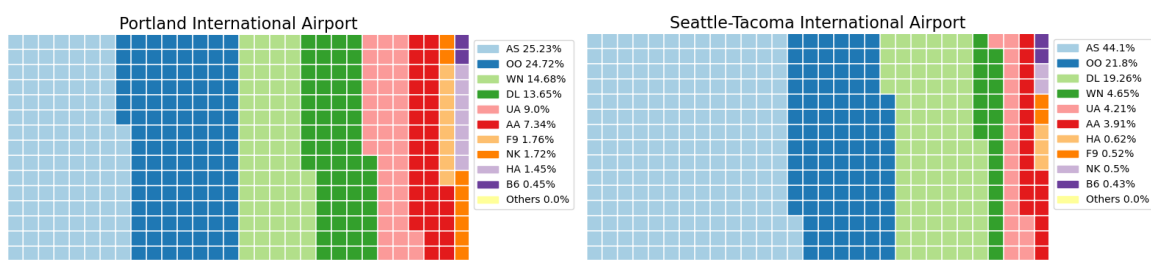

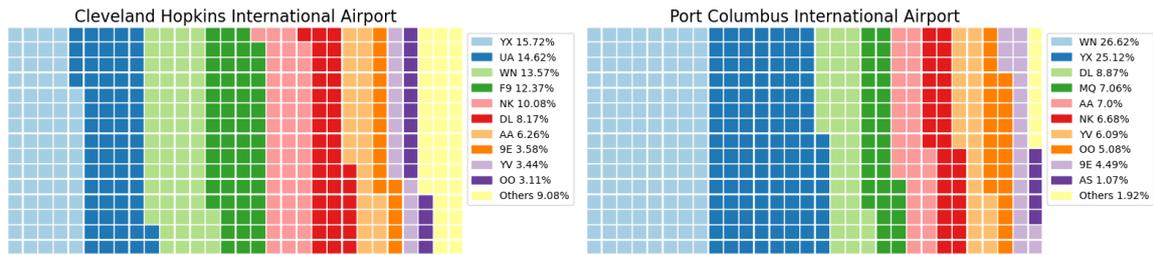

**Figure 9**. The share of flights of an airline at the other airports highly impacted the percentage change of dwell time per user. A high proportion of flights at these airports are operated by airlines applying a decentralized model, except for Seattle-Tacoma International Airport.

### 5.4. Impacts attributed to direct hazard exposure versus network effects

To understand to what extent impacts can be attributed to direct hazard exposure or network effects, we evaluated airport location and flight data. By plotting the impacts spatially in Figures 10(a) and 10(b), we can examine the extent to which the impacts can be attributed to direct hazard exposure (if airports with the greatest impacts are located in areas alerted by the National Weather Service). We can observe that airports located around the Great Lakes and the Northwest are the regions alerted by NWS and consistently had greater total dwell time and dwell time per user changes. The other airports also experienced significant changes in dwell time which may be due to indirect impacts like cascading network effects or strategically diverting the routes.

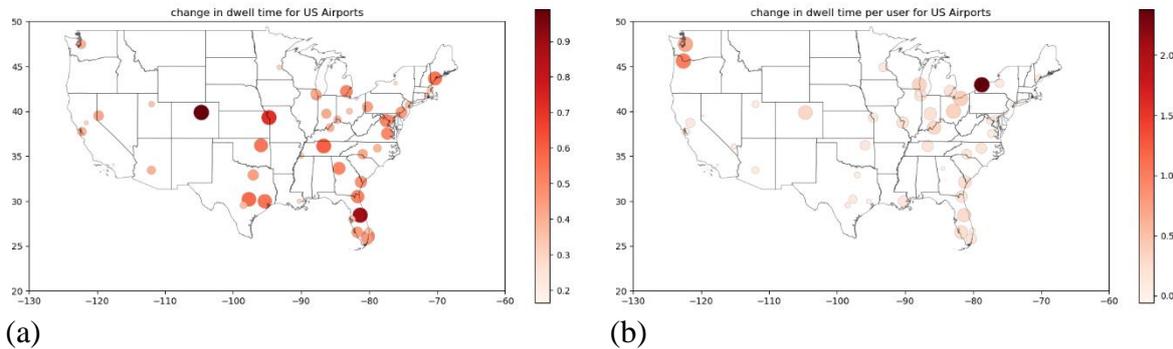

(a)                                                            (b)

**Figure 10**. Spatial distribution of percentage change in (a) total dwell time and (b) dwell time per user. Larger and darker markers indicate airports with greater changes in dwell time metrics. Airports located in the Great Lakes and Northwest were directly affected by the Winter Storm Elliot.

## 6. Discussion and Concluding Remarks

Extreme weather events have become more frequent and intense in recent years, causing significant disruptions to transportation systems, including air transportation systems. Winter Storm Elliott, which hit the United States in December 2022, was a prime example of such an event. Using fine-grained location-based datasets, this study investigated the effects of Winter Storm Elliott on the U.S. air transportation system, focusing on passenger delays as a metric of system performance.

The study offers multiple novel contributions. First, using observational data related to airport dwell time changes, the results quantified the overall impacts of the winter storm on the entire air transportation system in the United States from the perspective of passenger delays. Second, the study ranked airports most impacted by the Winter Storm and revealed the relationship between airline operation models and the vulnerability of airports to extreme weather events. Third, the use of location-based data in examining the vulnerability of air transportation systems shows the potential for adopting emerging datasets for better understanding of risks and vulnerability of air transportation systems to extreme weather hazards. Finally, departing from the majority of studies which focus on system performance (e.g., number of flight cancellations and average delay per flight), this study evaluated, with high resolution, the impacts of extreme events based on effects on passenger travel delays. The study's findings showed that Winter Storm Elliott had a significant impact on passenger dwell time at airports, resulting in significantly increased delays. The more vulnerable airports were located in areas prone to winter storms or with higher passenger flow. Interestingly, these vulnerable airports often shared a common characteristic: a large proportion of their flights were operated by airlines applying a decentralized operating model. The results suggest that such airports and airlines are more vulnerable to extreme weather events, which can result in a cascading effect on the rest of the aviation system.

In recent years, there has been a growing trend toward decentralization in various sectors, including the energy and transportation industries. For example, the concept of increasing energy security with a decentralized electric grid, also known as a microgrid, has gained significant traction. The power grid's decentralized architecture can improve its resilience to some extent, as power can be rerouted from unaffected areas to those affected by outages. Similarly, decentralized airline operations have also emerged as a potential solution to increase operational flexibility, reduce costs, and enhance customer experience. A decentralized airline operation can promote competition and reduce reliance on a few large airports, leading to more diverse and flexible flight options for passengers. However, decentralized airline operations can also be responsible for inefficiencies in passenger flow and aircraft utilization, as there is no centralized control to optimize flight schedules and coordinate operations. Decentralized airline operations can be more resilient under normal conditions or small disruptions, as delays or cancellations in one airport do not necessarily affect the entire network. Still, in extreme weather events, such an architecture can exacerbate the system's vulnerability, as delays in one airport can cascade to other airports, leading to system-wide disruptions.

In conclusion, this study highlights the need to improve the resilience of the air transportation system to extreme weather events. The results suggest that airports and airlines with a decentralized operation model are more vulnerable to such events, and strategies to mitigate their impacts should be developed. One possible strategy is to increase the system's redundancy, enabling it to absorb the impacts of extreme events more effectively. Additionally, policymakers and stakeholders should consider investing in infrastructure and technology to enhance the system's adaptability and resilience to such events. Overall, this study provides novel insights into the impacts of extreme weather events on the air transportation system and can inform future policies and strategies to improve resilience.

**Data availability**

All data were collected through a CCPA- and GDPR-compliant framework and utilized for research purposes. The data that support the findings of this study are available from Spectus, but restrictions apply to the availability of these data, which were used under license for the current study. The data can be accessed upon request submitted on Spectus.ai. Other data we used in this study are publicly available.